\documentclass[10pt,conference, onecolumn]{IEEEtran}
\usepackage{setspace}
 \usepackage{xargs}
 \usepackage{paralist}

\usepackage{graphicx}

\begin{document}

\title{Optimal Relay Placement For Capacity And Performance Improvement Using A Fluid Model For Heterogeneous Wireless Networks}


\author{\authorblockN{Jean-Marc Kelif,}
\authorblockA{Orange Labs\\
Issy-Les-Moulineaux, France\\
jeanmarc.kelif@orange-ftgroup.com}
\and
\authorblockN{Marceau Coupechoux and Marc Sigelle}
\authorblockA{T\'el\'ecom ParisTech and CNRS LTCI\\
46, rue Barrault, Paris, France\\
\{marceau.coupechoux,marc.sigelle\}@telecom-paristech.fr}
}

\maketitle

\begin{abstract}
In this paper, we address the problem of optimal relay placement in a cellular network assuming network densification, with the aim of maximizing cell capacity. In our model, a fraction of radio resources is dedicated to the base-station (BS)/relay nodes (RN) communication. In the remaining resources, BS and RN transmit simultaneously to users. During this phase, the network is {\it densified} in the sense that the transmitters density and so network capacity are increased. Intra- and inter-cell interference is taken into account in Signal to Interference plus Noise Ratio (SINR) simple formulas derived from a \textit{fluid model for heterogeneous network}. Optimization can then be quickly performed using Simulated Annealing. Performance results show that cell capacity is boosted thanks to densification despite a degradation of the signal quality. Bounds are also provided on the fraction of resources dedicated to the BS-RN link.   
\end{abstract}

%

\maketitle

\section{Introduction}

Relaying is a promising feature of future cellular networks. The scenarios envisioned by the two standards IEEE 802.16j (for WiMAX networks) and 3GPP Release 10 and 11 (LTE-Advanced) are the following: (a) coverage extension (including indoor coverage and shadowed zone mitigation), (b) capacity boost, (c) group mobility. In this paper, we tackle the problem of optimal relay placement for capacity increase in a cellular network. We assume simultaneous transmission of base stations (BS) and relay nodes (RN) so that the network is densified. We rely on simple formulas based on a fluid model in order to obtain quick results. Optimization is performed using Simulated Annealing (SA).
The relay placement problem arises in various contexts: wireless sensor networks, Wireless Local Area Networks (WLAN) and cellular networks. In this short literature review, we focus on the latter case. 


Authors of \cite{Lu09} consider the joint BS and RN placement problem for a given User Equipment (UE) distribution. Their objective is to maximize the system capacity with a pre-defined budget constraint. The resulting Integer Linear Progamming (ILP) problem is NP hard so that authors propose a two-stage deployment sub-optimal algorithm.  In our paper, we assume that traffic is uniform and we try to answer the question: how to place relays in a existing network in order to increase capacity ? The scope is thus narrower but we focus on optimal solutions.

Paper~\cite{Wang08} has similar objectives as ours. Here also, authors study the optimal relay placement problem assuming two server policies and two allocation strategies. The study has however two drawbacks. First, a single cell is considered, whereas in a real network, inter-cell interference plays a crucial role. In particular, relays placed at the cell edge experience the strong interference from RN and BS of neighboring cells. 
Secondly, authors do not take into account the densification (i.e., the increase of the transmitters density) induced by relays. If RN and BS are indeed able to transmit simultaneously, the cell capacity can be greatly improved \cite{KC09a}.

In \cite{Zhou11}, authors consider the joint optimization problem of RN placement and RN sleep/active probability in order to maximize energy efficiency in a cellular network. This work is very different from ours, the proposed model indeed assumes a linear network made of BS operating at different frequencies and does not allow inter-relay interference. Here, we assume that relays have no sleep mode and we account for interference in a single frequency hexagonal network.

A hierarchical optimization problem is formulated in \cite{Niyato09} for WiMAX networks: authors first focus on short term call admission control decisions (they use here the Markov Decision Process framework) and then on long term network planning (a binary ILP problem is solved with standard methods). Again, the interference calculation is out of the scope of the paper. Two other papers with different optimization objectives ignore interference: \cite{Yu08,Lin08}.

The organization of the paper is as follows. In Section~\ref{nwmodel}, we present the network model. In Sections~\ref{cellcapa} and \ref{sinreval}, we derive the cell capacity expression and related SINR formulas. Section~\ref{fluidmodel} recalls the principle of fluid model network approach and develops an extension to heterogeneous networks. Section~\ref{optim} describes the optimization algorithm. Some performance results are provided in Section~\ref{perfresults} and the last section concludes the paper.

\section{Network Model} \label{nwmodel}

In this section, we describe the considered network topology, the frames structure and the channel model.

\subsection{Network Topology}

We consider a single frequency cellular network consisting of omnidirectional eNodes-B (eNB) hexagonal cells. Let $R_c$ be the half-distance between two neighbor eNB and $\rho_{eN\!B}$ the eNB density. eNB transmit at power $P$.

\begin{figure}[htbp]
\begin{center}
\includegraphics[width=0.5\linewidth]{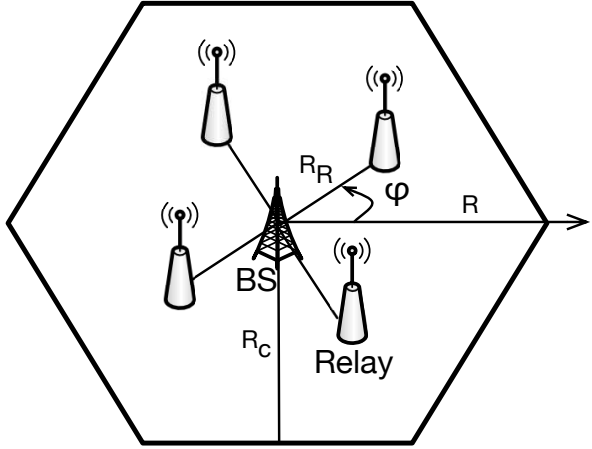}
\caption{Relay Deployment.}
\label{fig:relaydeployment}
\end{center}
\end{figure}

In each cell, $n$ RN are deployed with a regular pattern and controlled by the eNB. We focus on capacity evaluation for the downlink. The generic relay deployment is illustrated in Fig.~\ref{fig:relaydeployment}: relays are regularly deployed at distance $R_R$ from the eNB and at angles $\varphi+2\pi/n$, where $\varphi$ is an offset. Note that the deployment pattern is identical in all cells ($\varphi$ and $R_R$ are constant across the cells). The RN density is $\rho_R=n\rho_{eN\!B}$. RN transmit at power $P_R$.

RN in a cell are labeled from $1$ to $n$. Cells are labeled from $0$ to $B$ so that a RN can be uniquely identified by $(i,k)$, where $i$ is the relay number and $k$ is the cell number. A relay $(i,k)$ is said to be of {\it type i}. The set of type $i$ relays form a regular pattern, where the minimum half-distance between nodes is $R_c$ and whose density is $\rho_{eN\!B}$. 

We denote $r_b$ the distance between eNB $b$ and the UE of interest and, for simplicity, we set $r=r_0$. We denote $r_{i,k}$ the distance between the relay $(i,k)$ and the UE. Let also $r_{i,k^*(i)}=\min_{k}r_{i,k}$ be the minimum distance between the UE and a type {\it i} RN.

\subsection{Resource Organization}

We consider in-band half-duplex relays\footnote{The study can be extended to out of band full duplex relays, the following SINR evaluation is only slightly different.}. We assume a time division access between eNB and RN and we focus on a single frame of duration $t_{fr}=1$ (unit of time). The eNB transmits data to the relays over the Backhaul Link (BL) during a time $\tau$ and the eNB and the relays simultaneously transmit data to their respective attached UE during $1-\tau$, respectively over the Direct Link (DL, eNB-UE link) and the Relay Link (RL, RN-UE link).

We make the following assumption for the serving station selection: {\it best server}: the UE is served by the station from which it receives the highest power. 

We consider a scheduling scheme fair in radio resources: each transmitter (eNB or RN) allocates the same amount of radio resources to the UE attached to it (whatever its channel conditions). There is however no fairness between UE attached to eNB and UE attached to RN. Note that the frame structure is fixed whatever the proportion of UE attached to eNB or RN. 

%

\section{Cell Capacity} \label{cellcapa}

In this section, we define cell capacity and we consider two cases: constant and variable $\tau$. Cell capacity will be our criterion to compare different relay configurations.

\subsection{Fixed $\tau$}
In this section, we assume that $\tau$ is a constant. Let us concentrate on the central cell (cell with index $b=0$) and let $N_{tot}$, $p_{eNB}$, $p_{RNi}$ be resp. the number of UE in the cell, the proportions of UE attached to eNB and to RN $i$\footnote{We assume that there is a sufficient number of UE in the cell so that all nodes, eNB or RN, are active.}. Let $C_{eNB}$ and $C_{RNi}$ be the average capacities of resp. eNB and RN $i$:
\begin{eqnarray}
C_{eNB}&=&\frac{1}{N_{tot}p_{eNB}}\int_{\mathcal{S}_{eNB}}\rho(\mathbf{r})c(\mathbf{r})d\mathbf{r}, \\
C_{RNi}&=&\frac{1}{N_{tot}p_{RNi}}\int_{\mathcal{S}_{RNi}}\rho(\mathbf{r})c(\mathbf{r})d\mathbf{r},
\end{eqnarray}
where $\mathcal{S}_{eNB}$ (resp. $\mathcal{S}_{RNi}$) is the surface served by the eNB (resp. RN $i$), $\rho(\mathbf{r})$ is the UE density at location $\mathbf{r}$ and $c(\mathbf{r})$ is the throughput achievable at location $\mathbf{r}$. Throughout the paper, we will assume a constant UE density, so that $\rho(\mathbf{r})=\rho$. Parameter $c$ is a function of the SINR $\gamma$~\cite{Ellenbeck09}:
\begin{equation} \label{eq:approxshannon}
c(\mathbf{r})=
\left\{ \begin{array}{ll}
0 & \mbox{if } \gamma< -10\mbox{dB} \\
0.6W\log_2(1+\gamma(\mathbf{r})) & \mbox{if } -10\leq\gamma\leq 22\mbox{dB} \\
4.4 & \mbox{if } \gamma>22\mbox{dB},
\end{array}
\right.
\end{equation}
where $W$ is the system bandwidth. 

The amount of resources given to a UE attached to eNB is $(1-\tau)/(N_{tot}p_{eNB})$. The throughput for such a UE is $D_{eNB}=C_{eNB}(1-\tau)/(N_{tot}p_{eNB})$. RN $i$ controls $N_{RNi}$ UE. So the throughput achieved by a UE attached to RN $i$ is $D_{RNi}=C_{RNi}(1-\tau)/N_{RNi}$. The cell throughput when there are $n$ relays is thus given by:
\begin{eqnarray}
\nonumber C_{cell}&=&N_{tot}p_{eNB}D_{eNB}+\sum_{i=1}^{n}N_{tot}p_{RNi}D_{RNi}, \\
&=&(1-\tau)\left ( C_{eNB}+\sum_{i=1}^{n}C_{RNi}\right). \label{eq:ccell}
\end{eqnarray}

Let $C_{cell0}$ be the cell throughput when there is no relay ($\tau=0$ in this case). Relays bring an improvement iff $\tau\leq\tau^*(n)$ with:
\begin{equation}
\tau^*(n)=1-\frac{C_{cell0}}{C_{eNB}+\sum_{i=1}^{n}C_{RNi}}.
\end{equation}
Note that $\tau^*(n)$ is an increasing function of the total throughput delivered by relays.

\subsection{Variable $\tau$}

We now assume that $\tau$ is a function of the total throughput delivered by all the relays. In this case, we use the notation $\tau_B(n)$. Let $C_B$ be the backhaul throughput. The volume of data transmitted by RN $i$ in a frame is $C_{RNi}(1-\tau_B(n))$. This volume is transferred on the BL in $C_{RNi}(1-\tau_B(n))/C_B$ seconds. As a consequence, $\tau_B(n)$ verifies the following equation: 
\begin{eqnarray}
\nonumber \tau_B(n)&=&\sum_{i=1}^{n}\frac{C_{RNi}(1-\tau_B(n))}{C_B}.
\end{eqnarray}
And thus:
\begin{eqnarray}
\tau_B(n)&=&\frac{\sum_{i=1}^{n}\frac{C_{RNi}}{C_B}}{1+\sum_{i=1}^{n}\frac{C_{RNi}}{C_B}}.
\end{eqnarray}
With this assumption, the cell capacity definition is still given by (\ref{eq:ccell}) provided that $\tau$ is replaced by $\tau_B(n)$. Note that $\tau_B(n)$ is an increasing function of the total throughput delivered by relays.
%
%
%

\section{SINR Evaluation} \label{sinreval}

In this section, we evaluate the Signal to Interference plus Noise Ratio (SINR), $\gamma$, at a UE located at distance $r$ from the central eNB. We first assume that the UE is served on the DL by the eNB, then on the RL by a type $j$ RN. We further compute the capacity on the BL. Formulas are then simplified using the fluid model approach. 

\subsection{UE Served by the eNode-B}

Let $N_{th}$ be the thermal noise power. If the UE is attached to the eNB, UE is interferred by all the other eNB and all the relays of the network, we can thus write:
\begin{eqnarray}
\nonumber \gamma(r) &=& \frac{Pg(r)}{\displaystyle\sum_{b=1}^{B}Pg(r_b)+\displaystyle\sum_{k=0}^{B}\displaystyle\sum_{i=1}^{n}P_Rg_R(r_{i,k})+N_{th}} \\
\nonumber
&=&\frac{\gamma_0}{1+I_1+I_2}, \label{eq:gammaenb}
\end{eqnarray}
where
\begin{equation} \label{eq:gamma0}
\gamma_0=\frac{Pg(r)}{\displaystyle\sum_{b=1}^{B}Pg(r_b)},
\end{equation}
\begin{eqnarray}
\nonumber I_1&=&\frac{\displaystyle\sum_{i=1}^{n}\displaystyle\sum_{k=0}^{B}P_Rg_R(r_{i,k})}{\displaystyle\sum_{b=1}^{B}Pg(r_b)} \\
&=&\displaystyle\sum_{i=1}^{n}\Omega_i\left(1+\gamma_{i,k^*(i)}\right), \label{eq:i1}
\end{eqnarray}
\begin{eqnarray} \label{eq:gammaik}
\gamma_{i,k^*(i)}&=&\frac{P_Rg_R(r_{i,k^*(i)})}{\displaystyle\sum_{k=0, k\neq k^*(i)}^{B}P_Rg_R(r_{i,k})}, 
\end{eqnarray}
\begin{eqnarray} \label{eq:omegai}
\Omega_i&=&\frac{\displaystyle\sum_{k\neq k^*(i)}P_Rg_R(r_{i,k})}{\displaystyle\sum_{b=1}^{B}Pg(r_b)}, 
\end{eqnarray}
\begin{equation} \label{eq:i2}
I_2 = \frac{N_{th}}{\displaystyle\sum_{b=1}^{B}Pg(r_b)}.
\end{equation}
Parameter $\gamma_{0}$ can be interpreted as the SIR of the UE served the eNB $0$ if there were only eNB in the network (i.e., RN were not transmitting). In the same way, $\gamma_{i,k}$ can interpreted as the SIR of the UE if it were attached to relay $(i,k)$ in a network only composed of $i$-type relays (i.e., eNB and other types of RN are inactive). $\Omega_i$ is the ratio of interference received by type {\it i} relays and eNB. For the considered UE, relay $(i,k^*(i))$ is the closest of type {\it i}. Due to the network topology, it is possible that this relay is not controlled by the central eNB, so that $k^*(i)$ may be different from $0$.

\subsection{UE Served by a Relay Node}

If the UE is attached to a relay of type $j$, UE is interferred by all the eNB and all the other RN of the network, we can thus write:

\begin{equation}
\gamma(r_{j,k^*(j)})=\frac{\gamma_{j,k^*(j)}}{1+\frac{1+\gamma_0}{\Omega_j}+\displaystyle\sum_{i\neq j}(1+\gamma_{i,k^*(i)})\Omega_{i,j}+I_3}, \label{eq:gammarn}
\end{equation}

where
\begin{eqnarray} \label{eq:omegaij}
\Omega_{i,j}&=&\frac{\displaystyle\sum_{k\neq k^*(i)}P_Rg_R(r_{i,k})}{\displaystyle\sum_{k\neq k^*(j)}P_Rg_R(r_{j,k})}, 
\end{eqnarray}
\begin{eqnarray} \label{eq:i3}
I_3&=&\frac{N_{th}}{\displaystyle\sum_{k\neq k^*(j)}P_Rg_R(r_{j,k})}. 
\end{eqnarray}
Parameter $\Omega_{i,j}$ is the ratio of interference received by type {\it i} relays and type {\it j} relays.

\subsection{SINR on the Backhaul Link}

We consider now the backhaul link eNode-B-RN. The SINR, $\gamma_B$, on the BL at distance $R_R$ can be written as:
\begin{equation} \label{eq:gammab}
\gamma_B=\frac{Pg_B(R_R)}{\sum_{b=1}^{B}Pg_B(r_b)+N_{th}}
= \frac{1}{\frac{\sum_{b=1}^{B}Pg_B(r_b)}{Pg_B(R_R)}+I_4}
\end{equation}
\begin{eqnarray} \label{eq:i4}
I_4&=&\frac{N_{th}}{PK_BR_R^{-\eta_B}}.
\end{eqnarray}

\section{Fluid Model Networks}\label{fluidmodel}
\subsection{Fluid Model for Homogeneous Networks}
The fluid model~\cite{Kel05} is a powerful tool for simplifying SINR formulas in a wireless network. Consider a network of regularly spaced BSs with half inter-site distance equal to $R_c$, with density $\rho_{BS}$ and transmitting at the same power $P$. Let $g(r)=Kr^{-\eta}$ be the path-gain at distance $r$, where $K$ is a constant and $\eta$ is the path-loss exponent on the DL (in the same way, we introduce constants ($K_R$, $\eta_R$) for the RL). Assume that a UE is at distance $r$ from its serving BS. The total interference received by the UE can then be approximated by:
\begin{equation}
\frac{2\pi\rho_{BS}PK(2R_c-r)^{2-\eta}}{\eta-2}.
\end{equation}
We refer the reader to \cite{Kel10} for the detailed explanation and validation through Monte Carlo simulations. The main idea is to replace a discrete set of transmitters by a continuum and thus transform discrete sums into integrals. Beside its simplicity, the main advantage of this approach is to obtain a function that only depends on the distance to the serving BS rather than on all the distances to every interferer. We extend this concept to a cellular network with relays. 

\subsection{Fluid Model for Heterogeneous Networks}\label{fluidheterogeneous}
Contrary to what is assumed in \cite{Kel05}, such a network shows inhomogeneities: inter-distance between neighboring stations is not constant and BS and relay transmit powers are different. 

The network considered in this paper can however be seen as constituted of one regular subnetwork of BSs and $n$ regular subnetworks of relays. This is the basic idea of the extension of the fluid model to a cellular network with relays. In each of the relay subnetworks, as well as in the BS subnetwork, the half-inter-site distance is $R_c$. 
As a consequence, in our study, the fluid model can be used for computing the interference received from all BSs of the network on the one hand and from each type {\it i} ($i\in \{1,...,n \}$) relays on the other hand, taking into account all relays of the network. 
In line with the fluid model, we are able to simplify equations (\ref{eq:gamma0}), (\ref{eq:gammaik}), (\ref{eq:omegai}), (\ref{eq:i2}), (\ref{eq:omegaij}) and (\ref{eq:i3}):
\begin{eqnarray}
\gamma_0&=& \frac{(\eta-2)r^{-\eta}}{2\pi\rho_{BS}(2R_c-r)^{2-\eta}}\\
\gamma_{i,k^*(i)}&=& \frac{r_{i,k^*(i)}^{-\eta_R}(\eta_R-2)}{2\pi \rho_{BS}(2R_c-r_{i,k^*(i)})^{2-\eta_R}} \\
\Omega_i&=& \frac{P_RK_R(2R_c-r_{i,k^*(i)})^{2-\eta_R}(\eta-2)}{PK(2R_c-r)^{2-\eta}(\eta_R-2)} \\
I_2&=& \frac{N_{th}(\eta-2)}{2\pi \rho_{BS}PK(2R_c-r)^{2-\eta}} \\
\Omega_{i,j}&=& \frac{(2R_c-r_{i,k^*(i)})^{2-\eta_R}}{(2R_c-r_{j,k^*(j)})^{2-\eta_R}} \\
I_3&=& \frac{N_{th}(\eta_R-2)}{2\pi \rho_{BS}P_RK_R(2R_c-r_{j,k^*(j)})^{2-\eta_R}} 
\end{eqnarray}

Above equations, along with equations (\ref{eq:gammaenb}), (\ref{eq:i1}) and (\ref{eq:gammarn}), allow to quickly compute the SINR of a terminal in the cell of interest. The only required distances for this computation are the distance to the BS and the distance for each {\it i} to the nearest type {\it i} relay. 

\subsection{Validation of the Fluid Model for Heterogeneous Networks} \label{fluidvalid}

In this section, we propose a validation of the \textit{Fluid Model for Heterogeneous Networks} presented in the last section. In this perspective, we will compare the Cumulative Distribution Function (CDF) of the SINR obtained by using fluid expressions established in section \ref{fluidheterogeneous} to those obtained numerically by Monte Carlo simulations. Our simulator assumes a central hexagonal cell surrounded by 10 rings of interferers. Moreover, 3 RNs are located in each macro cell. The distance  between a eNB and its associated RNs is $R_R$, the transmitting powers are respectively 31 dBm (RNs) and 43 dBm (eNBs). We assume a uniform distribution of UEs.

\begin{figure}[htbp]
\centering
\includegraphics[width=0.7\linewidth]{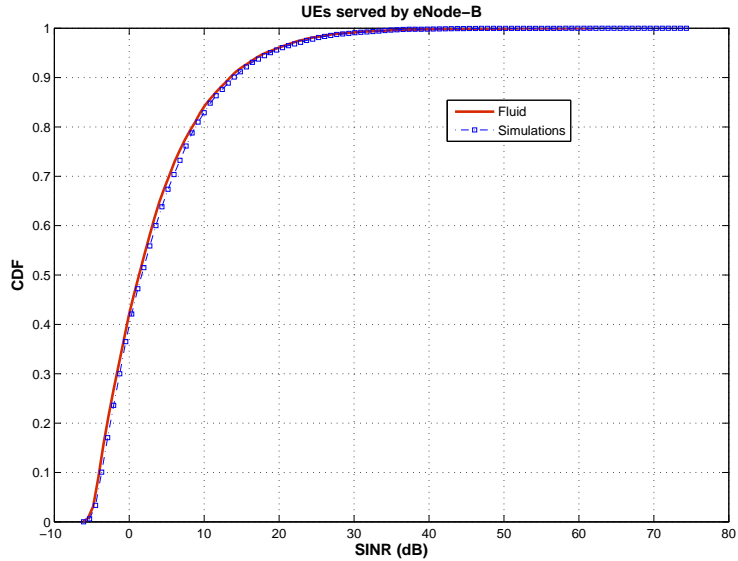}
\caption{\footnotesize CDF of the SINR for a UE connected to a eNB for $R_R=0.7 R_c$.}
\label{Validation}
\end{figure}

\begin{figure}[htbp]
\centering
\includegraphics[width=0.7\linewidth]{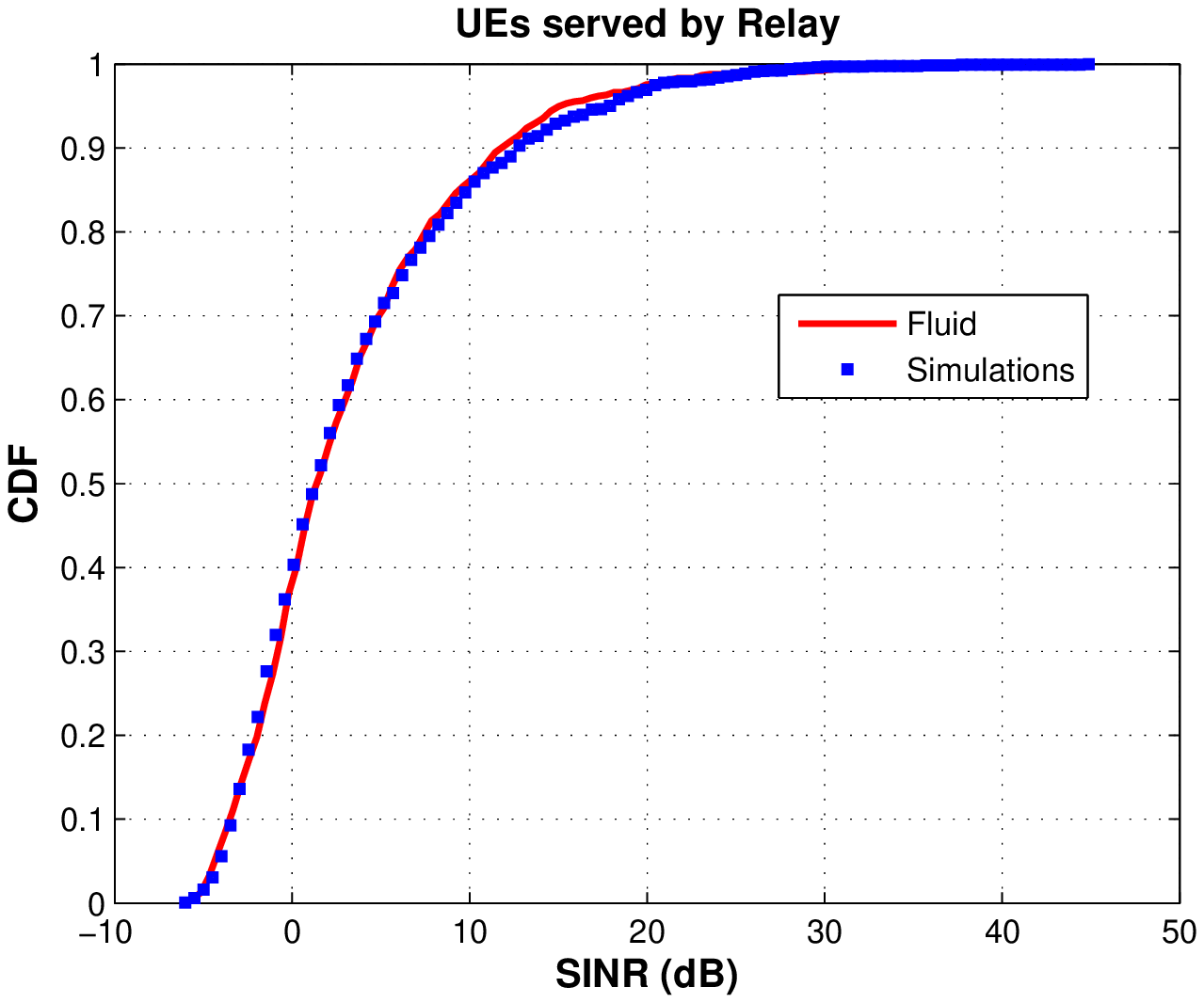}
\caption{\footnotesize CDF of the SINR for a UE connected to a RN for $R_R=0.7 R_c$.}
\label{Validation2}
\end{figure}

\begin{figure}[htbp]
\centering
\includegraphics[width=0.7\linewidth]{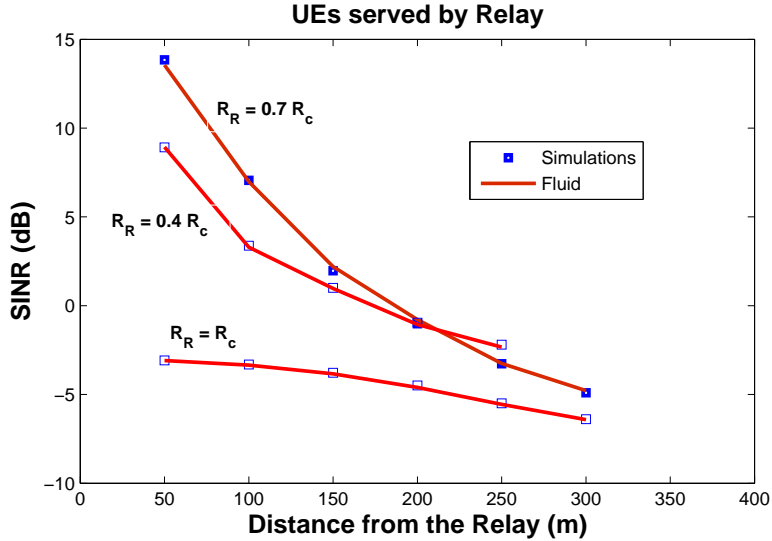}
\caption{\footnotesize Average SINR for a UE connected to a RN for $R_R=0.4 R_c$,   $R_R=0.7 R_c$ and $R_R=R_c$.}
\label{Validation3}
\end{figure}

Figures \ref{Validation} and \ref{Validation2}  show that the CDF calculated by using the fluid model for relay-based networks are very close to the ones obtained by simulations. Figure  \ref{Validation3} shows the variation of the average UE SINR connected to a relay with respect to the distance to this relay. It can be observed that the SINR established by simulations are the same than the ones calculated by the fluid model. The SINR is a decreasing function of $R_R/R_c$ because useful received power decreases, and interference increases, with the distance to the serving station.

\section{Optimal Configurations} \label{optim}

In this section, we look for the optimal relay configurations in terms of cell capacity: $\max C_{cell}$, where $C_{cell}$ is given by (\ref{eq:ccell}) and the optimization is done w.r.t the following variables: 
\begin{itemize}
	\item the number of relays: $n=0, 1,..., 6$;
	\item the eNB-relay distance: $R_R\in [0;R_c]$; 
	\item the offset angle: $\varphi\in [0;\pi/2]$;
	\item the relay transmit power: $P_R\in [18;31]$~dBm.
\end{itemize}
Respective steps are $\Delta n=1$, $\Delta R_R=0.1R_c$, $\Delta \varphi=\pi/20$, $\Delta P_R=1$~dBm.

A state is a combination of these variables. As the number of possible configurations is very large, we have to rely on an optimization technique. For example, in this paper, we use Simulated Annealing  
for its simplicity and because it statistically guarantees to find an optimal solution. 
\newcommand{\bbbr}{{\rm I\!R}}
Indeed, SA is a well-known stochastic technique
for solving large (but finite) combinatorial optimization problems.
It amounts back to the fifties
\cite{Metropolis1953}
but was rediscovered in the eighties 
\cite{Kirkpatrick1983,Laarhoven1987},
with successful applications 
to network optimization 
\cite{Bonomi1984,Keung2010}.

Its principle is the following:
let $\Omega$  be a {\bf finite} configuration space
and consider a cost (energy) function 
$U(x): \Omega \mapsto \bbbr$.
A minimizer of $U(.)$ can be found as follows:
\begin{itemize}
\item Start from any configuration $x_0 \in \Omega$ at step $0$.
\item Iteratively repeat the following process: Let $x_m$ be the current configuration at step $m$. 
Then draw a candidate configuration $\xi  \in \Omega$
at random and compute the associated energy variation
$\Delta U = U(\xi) - U(x_m)$.
If $\Delta U \leq 0 $ then assign $x_{m+1} = \xi$.
Else, assign $x_{m+1} = \xi$ with probability 
$p = \exp - \displaystyle \frac{\Delta U}{T_m} $
and  $x_{m+1} = x_m$ with probability 
$1-p$\enspace.
\end{itemize}

Here $T_m$ is an external {\em positive} {\em temperature} parameter
such that 
$ \displaystyle \lim_{m \rightarrow+\infty} T_m = 0^+$. Temperature is initialized at $T_0=35$ and is updated 
at step $m \geq 1$ as: $T_m =\alpha T_{m-1}$,
where $\alpha=0.995$ is a constant. According to the considered performance parameter, energy of a candidate is the opposite of the cell capacity: $U=-C_{cell}$. The number of iterations is $2000$ in our experiments. Thanks to the simplified formulas furnished by the fluid model, finding an optimal solution for a set of parameters is relatively quick (around $100$~s on a average laptop). 

\section{Performance Results} \label{perfresults}

In this section, we provide some performance results assuming the following set of parameters
~\cite{36.814} : $\eta=4.28$,$\eta_R=3.75$, $K=1.86$, $K_R=1.9e+3$, $R=1$~Km, $N_{th}=-104$~dBm, $W=10$~MHz and $P=43$~dBm. Each UE selects its serving node according to the best server policy.

\subsection{Optimal locations of relays}
\begin{figure}[htbp]
\begin{center}
\includegraphics[width=\linewidth]{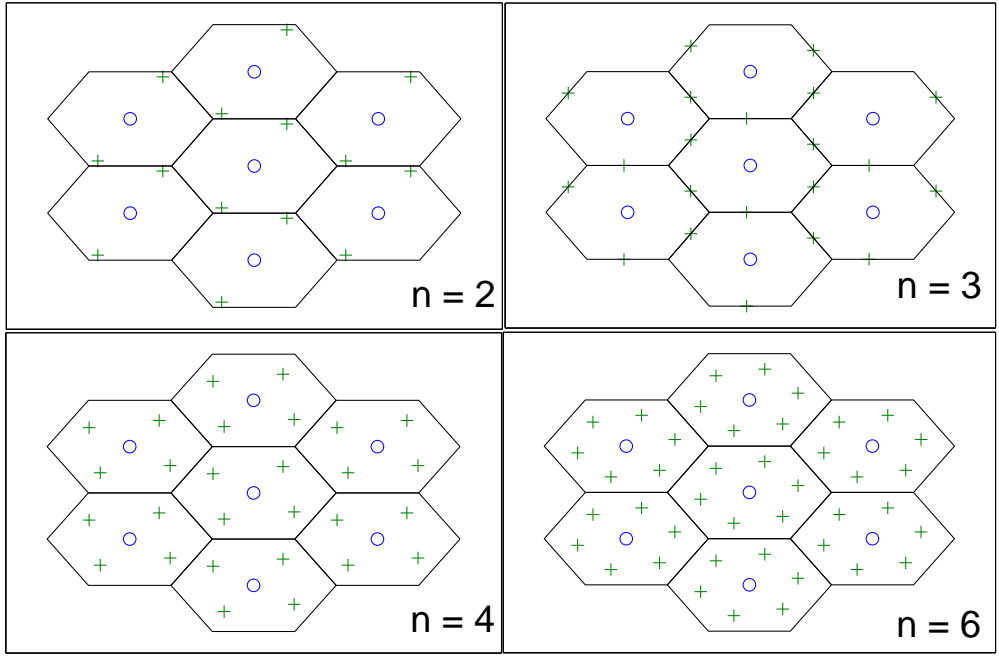}
\caption{Optimal RS placements for 2, 3, 4 and 6 relays per cell and assuming best server selection.}
\label{fig:configs}
\end{center}
\end{figure}

For resp. $n=\{1,2,3,4,5,6\}$ relays and best server selection criterion, optimal solutions are given by $\varphi=\{\frac{7\pi}{20},\frac{7\pi}{20},\frac{3\pi}{20},\frac{6\pi}{20},\frac{9\pi}{20},\frac{8\pi}{20}\}$, $R_R=\{1,1,1,0.7,0.7,0.7\}*R_c$ and $P_R=\{29,18,20,25,25,20\}$~dBm.
Fig.~\ref{fig:configs} shows some examples of optimal configurations with 2 to 6 relays in this case. Two kind of optimal configurations can be observed according to the distances between relays and eNB: configurations where $R_R = R_c$ and configurations where $R_R = 0.7 R_c$. The first one corresponds to a location of relays at eNB cell edges. They appear for low numbers of relays, n={1,2,3}. The second one corresponds to a location of relays at eNB cell centers. They appear for relatively high number of relays, n={4,5,6}. It is interesting to observe that none of these configurations uses the maximal available relay transmit power (set to 31 dBm, in our analysis). Except for the case n=1 for which the transmit power has a relatively high level (29 dBm), all the optimal configurations need low relay transmit power, comprised between 18 and 25 dBm.  It means that it relay don't need to transmit with a high power to maximize performances. 


\subsection{Optimal Performances}
\begin{figure}[htbp]
\begin{center}
\includegraphics[width=\linewidth]{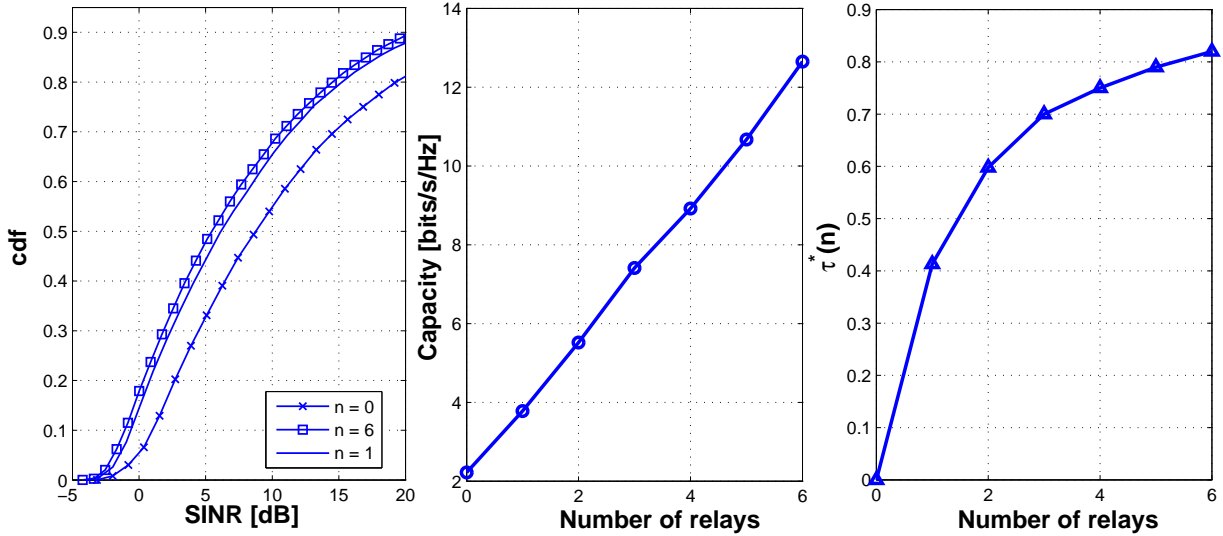}
\caption{SINR distribution as a function of the number of relays at their optimal locations (assuming best server selection).}
\label{fig:sinrcdf}
\end{center}
\end{figure}

Fig.~\ref{fig:sinrcdf} shows the effect of densification on the SINR CDF when best server selection is assumed. The SINR CDF allows to characterizes the performances distribution and the outage probability, too. As the number of interfering stations increases with the number of RN and as we supposed that relays and eNB transmit simultaneously, UE experience a lower signal quality. The presence of relays in the system has a significant impact on the CDF. For example, it can be observed a loss of about 3 dB, for an outage of 10\% when there are relays.  It is very likely that locally, some UE see their SINR improved with relays, however, we observe a global degradation of the SINR. Note that worst UE do not see their signal quality decrease (in the low SINR values, CDF almost coincide). However, when the number of relays increases, the difference between CDFs with and without relays does not increase.


\begin{figure}[htbp]
\begin{center}
\includegraphics[width=0.8\linewidth]{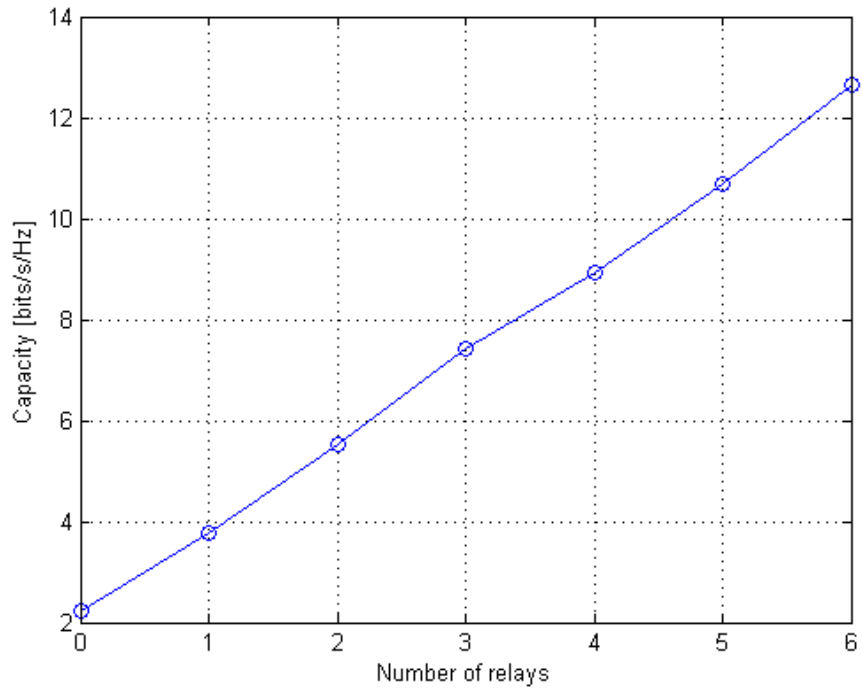}
\caption{Spectral efficiency per cell as a function of the number of relays (at their optimal locations) assuming best server$\tau=0$.}
\label{fig:capacity}
\end{center}
\end{figure}

The signal quality degradation is however compensated by the increase cell capacity. Fig.~\ref{fig:capacity} shows how the spectral efficiency per cell is increasing with the number of relays. This is indeed due to the fact that, within a cell, several UE are simultaneously served (by eNB and RN). With our assumption on the frame structure, relay deployment (with $\tau=0$) is equivalent to a network densification, which is known to increase the network capacity (see \cite{KC09a}).

\begin{figure}[htbp]
\begin{center}
\includegraphics[width=0.8\linewidth]{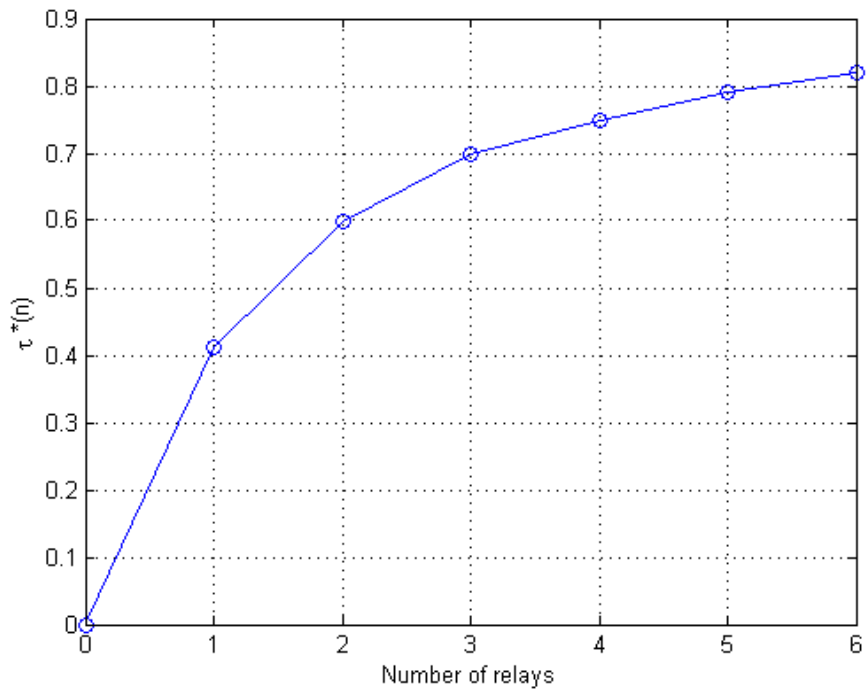}
\caption{Threshold value for $\tau$ as a function of the number of relays ($\tau^*(n)$) at their optimal locations.}
\label{fig:taustar}
\end{center}
\end{figure}

We now consider the loss due to the transmission on the BL, while taking into account parameter $\tau$ (constant). Fig.~\ref{fig:taustar} gives the threshold value $\tau^*(n)$ below which it is interesting to deploy relays. As an illustrative example, with the best server selection, it is worth deploying four relays provided that the BL does not consume more than 80\% of the radio resources.

If now $\tau=\tau_B(n)$ is an increasing function of the number of relays, increasing $n$ leads to capacity increase through network densification, it however also increases the amount of resources dedicated to the backhaul link. We assume that the BL benefits from a very good quality and let $C_B=4.4$~bits/s/Hz\footnote{$C_B$ can be alternatively computed using (\ref{eq:approxshannon}), we however leave this for further study.} and we use SA. The optimal solution for best server selection is $n=6$, $\varphi=\frac{\pi}{10}$, $P_R=18$~dBm, and $R_R=0.7R_c$; with these results, we can compute $\tau_B(n)=0.70$. 
We conclude that the gain brought by densification compensates more than the loss due to BL resources and that it is always advantageous to increase the number of relays.

\subsection{Impact of propagation} 


It appears interesting to analyze the impact of the propagation on their optimal location. In this aim, we set $P_R$= 20 dBm $P_B$=46dBm $K = 1.86$ and we vary $K_R$. Different parameters may induce a variation of this last parameter such as the environment, the antenna, the high of antenna. We analyze the case of 6 relays.

\begin{figure}[htbp]
\begin{center}
\includegraphics[width=0.8\linewidth]{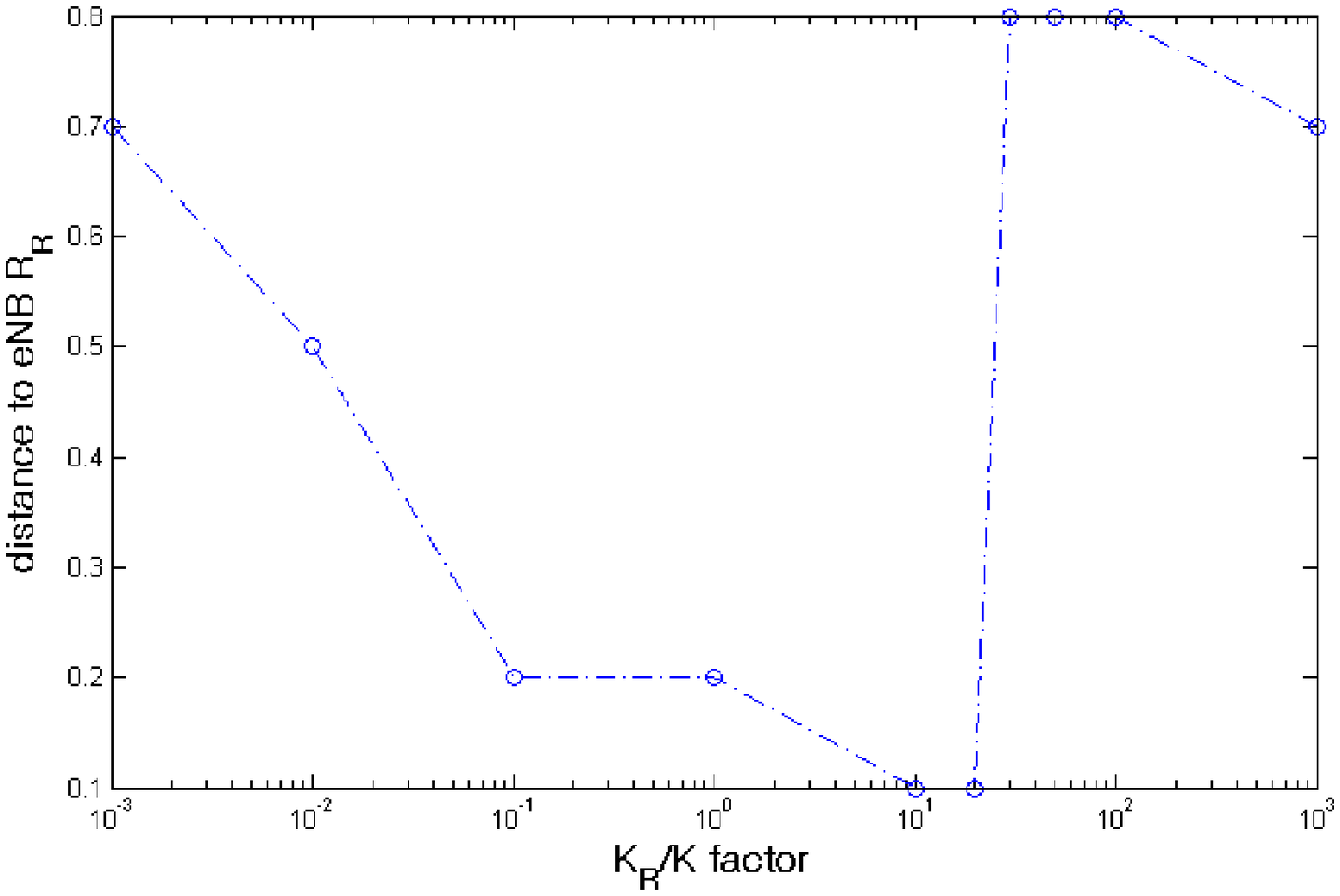}
\caption{Optimal location of relays $R_R$ as a function of $\omega_R = K_R/K$ and for n=6 relays.}
\label{fig:RRvsKR}
\end{center}
\end{figure}

\begin{figure}[htbp]
\begin{center}
\includegraphics[width=0.8\linewidth]{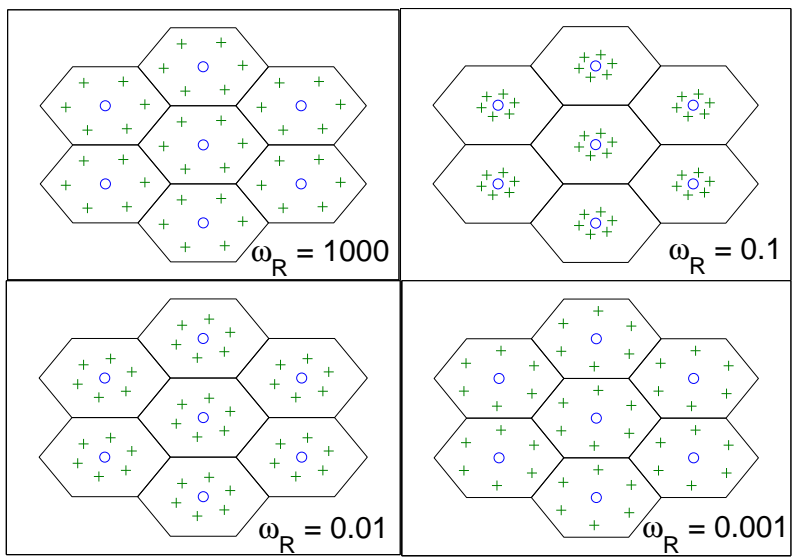}
\caption{Optimal location of relays for $\omega_R =$ 1000, 0.1, 0.01, 0.001 and for n=6 relays.}
\label{fig:configvsK}
\end{center}
\end{figure}

Denoting $\omega_R = \frac{K_R}{K}$, 
fig.~\ref{fig:RRvsKR} and \ref{fig:configvsK} show the variations of the optimal location of relays according to the variations of $\omega_R$.
We observe three zones. 
For high values of $\omega_R$, which correpond to the case $K_R >> K$, optimal locations of relays are far from eNBs. These locations answer to a need of limitation of interferences between eNB and relays. For $K_R \approx K$, powers received from relays are relatively low and consequently $R_R$ optimal may be close to the eNB, without generating important level of interferences. For low values of $\omega_R$, which correponds to the case $K_R << K$, the impact of thermal noise $N_{th}$ becomes very sensitive : this impact is mostly far from eNB. Consequently, relays have to be far from eNB to compensate this impact. Optimal locations of relays are thus far from eNBs. These locations correspond to a coverage need since relays allow to improve the coverage of eNB. 

\section{Conclusion}
In this paper, we have studied the problem of optimal RN placement in a cellular network, while taking into account the whole interference created by the network. Contrary to many papers, we have optimized the cell capacity by allowing relays and eNodes-B to transmit simultaneously. We have found the optimal RN locations and transmit powers using Simulated Annealing. The simple formulas provided by the fluid model allow a quick finding of the optimal solutions. We have shown that our approach tends to degrade the overall signal quality. Some radio resources should also be dedicated to the backhaul link between eNode-B and relays. However, the gain achieved by network densification largely outperforms the observed losses. Our further work includes the study of shadowing impact in this context. Other selection criteria are also foreseen. We at last intend to study inhomogeneous relay deployments across adjacent cells. 

\bibliographystyle{unsrt} 
\bibliography{bibliorelays,mysa} 

\end{document}